\documentclass[preprint]{aastex}

\received{}
\accepted{}
\journalid{}{}
\articleid{}{}
\slugcomment{}

\shortauthors{Budav\'ari {\it et~al.}}
\shorttitle{Multihump filter}
\newcommand{\etal}{{\it et~al.}}

\begin{document}
\title{Optimal multihump filter for photometric redshifts}

\author{Tam\'as  Budav\'ari\altaffilmark{1}, Alexander S.  Szalay}
\affil{Department of Physics and   Astronomy, The Johns  Hopkins  University,
Baltimore, MD~21218}
\email{budavari@pha.jhu.edu}
\author{Istv\'an  Csabai\altaffilmark{2}}
\affil{Department of Physics, E\"{o}tv\"{o}s University, Budapest,
Pf.\ 32, Hungary, H-1518}
\author{Andrew J. Connolly}
\affil{Department of Physics and Astronomy, University of
Pittsburgh, Pittsburgh, PA~15260}
\and
\author{Zlatan Tsvetanov}
\affil{Department of Physics and   Astronomy, The Johns  Hopkins  University,
Baltimore, MD~21218}

\altaffiltext{1}{Department of Physics, E\"{o}tv\"{o}s University, Budapest,
Pf.\ 32, Hungary, H-1518} 
\altaffiltext{2}{Department of Physics and   Astronomy, The Johns  Hopkins
University, Baltimore, MD~21218}

\begin{abstract}
We propose a novel type filter for multicolor imaging to improve on the
photometric redshift estimation of galaxies.  An extra filter---specific to a
certain photometric system---may be utilized with high efficiency.  We
present a case study of the {\it Hubble Space Telescope}'s Advanced Camera
for Surveys and show that one extra exposure could cut down the mean square
error on photometric redshifts by 34\% over the $z<1.3$ redshift range.
\end{abstract}

\keywords{galaxies: distances and redshifts --- galaxies: photometry --- 
instrumentation: miscellaneous}

\section{Introduction} \label{sec:multi.intro}

The  prediction of photometric  redshifts \citep{koo85,  connolly95a, gwyn96,
sawicki97,  hogg98, wang98,  yee98, soto99,  weymann99,  benitez00, csabai00,
budavari00} is built upon global  continuum features of the underlying galaxy
spectrum such as  the Balmer break at 4000\,{\AA}.  The more precisely
one can  localize these  features, the more  accurate the  redshift estimates
become.  However, traditional broadband  filters were not designed to provide
the best  possible photometric redshifts, for several reasons.  In  theory, a
set of filters optimal for photometric redshift estimation might be designed,
but this would probably be substantially different from any current standard
set.   Here we  do not  intend to  solve the  generic problem  of  finding an
optimal filter set for photometric  redshift estimation; this study shows how
conventional  filter  sets  can  be  extended  to  yield  better  photometric
redshifts.

In Section~\ref{sec:multi.idea}, we present the idea of an additional filter
and look into why and how an extra measurement---sampling the same
wavelengths---can improve significantly on photometric redshifts.  In
Section~\ref{sec:multi.design}, the algorithm is developed and we make a case
study of the Advanced Camera for Surveys of the {\it Hubble Space Telescope}.
Based on our analysis, we propose a novel type of ``broad-band'' filter that
can be effectively used for photometric redshifts.

\subsection{The idea} \label{sec:multi.idea}

The more general inversion problem of photometry in terms of redshift, type,
and intrinsic luminosity \citep{koo85, connolly95a, connolly95b, connolly99,
budavari99, budavari00, csabai00} involves the study of the Jacobian matrix
of fluxes as a function of the physical parameters of galaxies.  Our
numerical simulations show that filter sets yield more precise estimates when
continuum spectral features are located around the overlap of two
bands---both fluxes change rapidly with redshift.  One way to improve the
redshift estimates is to increase the number of filters and use narrower
bands.  This would require too many exposures and yield a non-standard
photometric system \citep{hickson98}.  The other possibility is to put an
intermediate width (IM) filter in a broad band, which would help the redshift
estimation by improving the resolution.  A single IM band can only help in a
narrow redshift range, so additional, narrower filters are needed for each
broad band.  This approach could triple the resolution but will heavily
increase the total exposure time at the same signal-to-noise ratio.

We propose combining multiple IM filters located in standard photometric
bands into one ``broadband'' filter.  We call this a {\it multihump} filter.
The resulting new filter will be able to improve the redshift prediction
(wavelength resolution) over a much wider redshift range and will only
require a single additional exposure of comparable length.

How does this really work?  Our approach makes a large difference for blue
galaxies.  The 4000\,{\AA} break in their spectra shows up only as a small
``bump'' which makes the redshift estimation difficult.  Let us consider the
worst-case scenario, a toy model spectrum of a late-type galaxy that declines
with wavelength as a power law and has as its only spectral feature the
approximately 500\,{\AA} wide bump.  Without this feature, the colors would
be the same at any redshift.  The flux changes with redshift within a band
depending on the shape of the filter.  If its gradient is too small (e.g.\
flat as a top-hat), tiny photometric errors can yield large errors in the
redshift prediction.  In other words, while the bump is in a broad band, it
is hard to tell precisely where it is.  The measured flux in our new filter
will change in redshift regions where the spectral feature is passing through
one of the IM humps.  These are the regions where the proposed filter
contributes the most to the redshift estimation and where the original filter
set has poor performance (small gradient).  Red elliptical (E/S0) galaxies do
not present a significant problem, because the strong discontinuity at
4000\,{\AA} changes noticeably the fluxes as it passes through the bands with
redshift.  In this case the additional filter plays a subsidiary role, and
the improvement of the accuracy is incremental.

\section{Filter design} \label{sec:multi.design}

The  actual  throughput of  a  photometric  system  strongly depends  on  the
instrument, different  projects use different CCDs to  gain more sensitivity
in the desired wavelength range.  The quantum efficiency of these devices set
strict constraints  on the  observable wavelength regime.  The filter  set is
selected to cover the available spectral interval according to science goals.
The multihump filter  is specific to a filter set, and its transmissivity is
affected by the CCD's quantum efficiency (QE).

We have analyzed the photometric system of the {\it Hubble Space Telescope}'s
Advanced Camera for Surveys
\citep[ACS][]{ford96}\footnote{http://adcam.pha.jhu.edu}.  The two
back-illuminated CCDs of the wide field channel are VIS-AR coated in
order to optimize the response in the $I$ band, and thus the $U$-band sensitivity
has been sacrificed.  Mission specifications dictate the constraints on the
multihump filter that is to extend the original set of the {\it g', r', i'}
and {\it z'} bands, which span the entire observable wavelength range. No bluer
or redder filter can be used, because the detector is not sensitive at shorter
or longer wavelengths.

The framework of the algorithm is simple. We parameterize a fictitious filter
curve and  convolve it with the CCD's  actual QE to obtain a realistic response
function.  Simulated photometric catalogs  of fluxes and errors are generated
using the  extended filter set.  The  goodness of a particular  filter set is
assessed by computing the errors on the redshift estimates.

\begin{figure}
\plotone{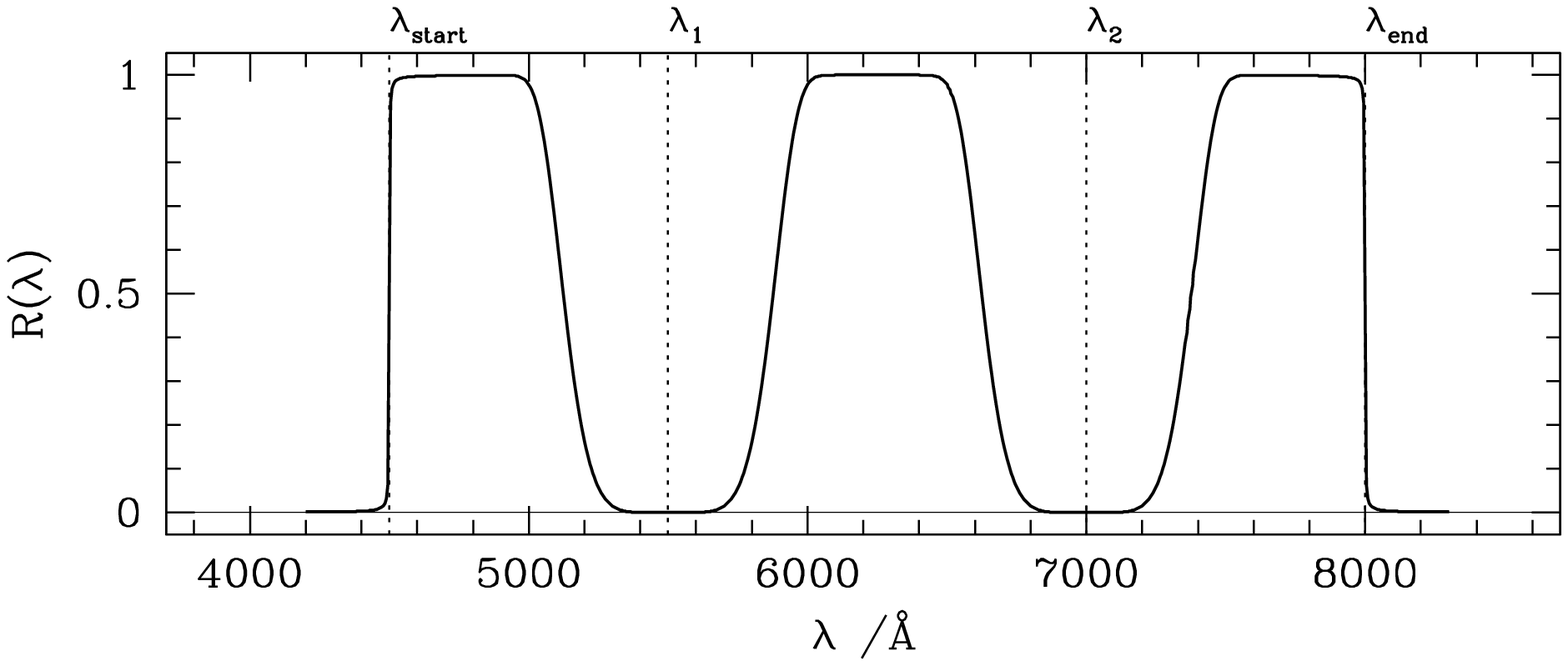}
\caption[Filter's response curve]{Illustration of  the filter's response
curve, as  defined by Equation~\ref{eq:multi.eqraw}.} 
\label{fig:multi.figraw}
\end{figure}

In the context of current filter manufacturing techniques \citep{offer98}, a
feasible parameterization of the additional filter's response function
($R(\lambda)$, see Figure~\ref{fig:multi.figraw}) consists of at most two
notch filters, plus a low and a high cutoff, as described by 
\begin{eqnarray}
R(\lambda) & = & 
{1\over \pi} \arctan \Big(\epsilon \cdot (\lambda -\lambda_{\rm start}) \Big) 
- {1\over \pi} \arctan \Big(\epsilon \cdot (\lambda -\lambda_{\rm end}) \Big) 
\nonumber \\ 
& & - \exp 
\Big[ - \Big( {\lambda - \lambda_1 \over \Delta\lambda_1} \Big)^p \Big] 
    - \exp 
\Big[ - \Big( {\lambda - \lambda_2 \over \Delta\lambda_2} \Big)^p \Big].
\label{eq:multi.eqraw}
\end{eqnarray}
Parameters $\lambda_{\rm  start}$ and $\lambda_{\rm  end}$ are the  limits of
the filter, and  $\epsilon$ determines the gradients, the  rise of the edges.
The values of  $\lambda_1$ and $\lambda_2$ correspond to  the locations of the
notches and $p$ (an even number) changes their shape. 

The mock catalogs were generated based on spectral energy distributions
(SEDs) derived from a physical interpolation scheme applied to the spectra of
\citet{cww}, as described in \citet{connolly99} and \citet{budavari00}. On a
high resolution redshift grid ($\delta z=5 \times 10^{-3}$), a hundred
objects were generated at each grid point with random SEDs.  The continuous
type parameter was scaled to yield a population of galaxies in which the
probability of having a spectral type between Ell and Sbc was 10\%, Sbc--Scd
35\%, Scd--Irr 35\%, and bluer 20\%.  The fluxes were modulated with 3\%
photometric errors taken from a Gaussian distribution.  The quality of the
photometric redshifts is determined by computing the rms of the estimates.
The error is calculated for the desired redshift interval (or regions)
meeting the actual project requirements---in this study, redshifts between
$z=0.2$ and $z=1.2$.

The actual optimization problem has quite a few parameters (see
Equation~\ref{eq:multi.eqraw}) to solve for, which makes it hard to
numerically find the global minimum.  It is also quite time-consuming to
evaluate the underlying function; a new loop of the simulation has to be
completed each time. Also extra constraints were added to the problem.  To be
able to use the very same filter from the ground for testing purposes or in
follow-up observations, a design compromise due to the atmosphere has been
made. In the actual optimization algorithm, one of the notches was always
forced to include the 5577\,{\AA} sky line.  To map the entire parameter
space, a linear search on a coarse grid was used to find the best possible
filter candidates, and then from the those minima, a gradient method was
utilized with discrete, higher resolution step-sizes.

\section{Discussion} \label{sec:multi.disc}

\begin{figure}
\plotone{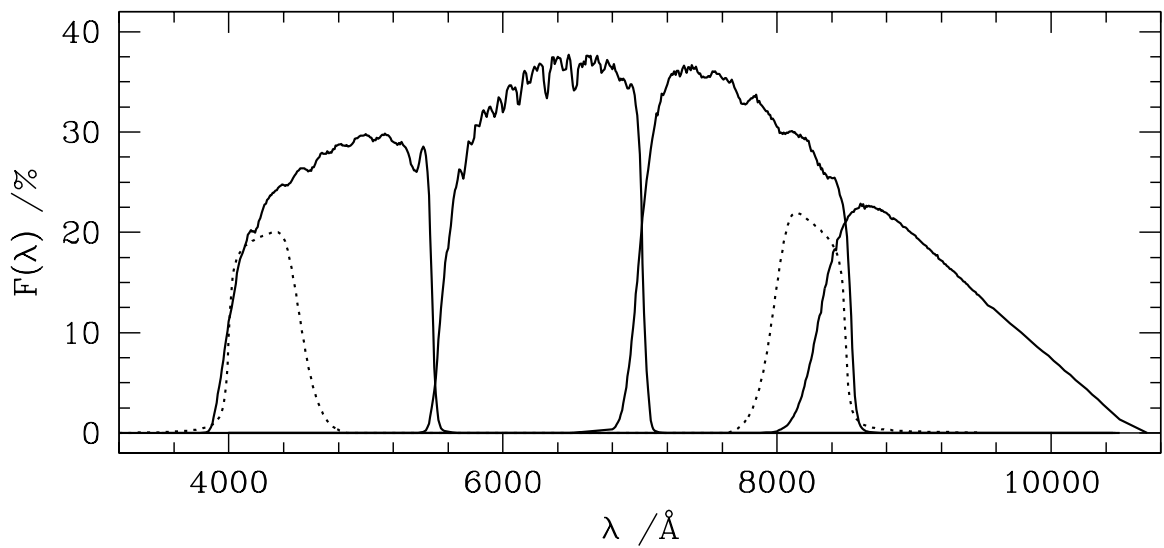}
\caption[Best extended filter set]{The throughput of the filters as used in
convolutions. The original {\it g',r',i',} and {\it z'} bands are represented
by solid lines ({\it left to right}).  The add-on multihump filter ({\it dotted
line}) is scaled down by 30\% for illustration.}
\label{fig:multi.figfilt}
\end{figure}

The resulting shape of the multihump is shown in
Figure~\ref{fig:multi.figfilt}. This best configuration has two humps and its
transmissivity peaks at the shortest possible wavelengths in the {\it g'}
band and, at longer wavelength, in the {\it i'} band. Since the ACS filters
look almost like triangles (as a result of the strong wavelength dependency in the
CCD's QE), our additional filter tries to intensify the changes in colors, at
the lowest, $z \sim 0.2$, and highest, $z \sim 1.2$, redshifts.  On one side,
the optimal multihump filter is essentially trying to compensate the lack of
a ultraviolet filter by boosting the contrast at the edge of the {\it g'}
band.  In fact, the algorithm also finds an optimal solution of having a $U$
band, which is not a real option because of the CCD's QE (see
Section~\ref{sec:multi.design}).  The location of the hump in the {\it i'} is
clearly set by the upper redshift limit $z \lesssim 1.2$, when the Balmer
break is at~$\lesssim 8800$\,{\AA}.

\begin{figure}
\plotone{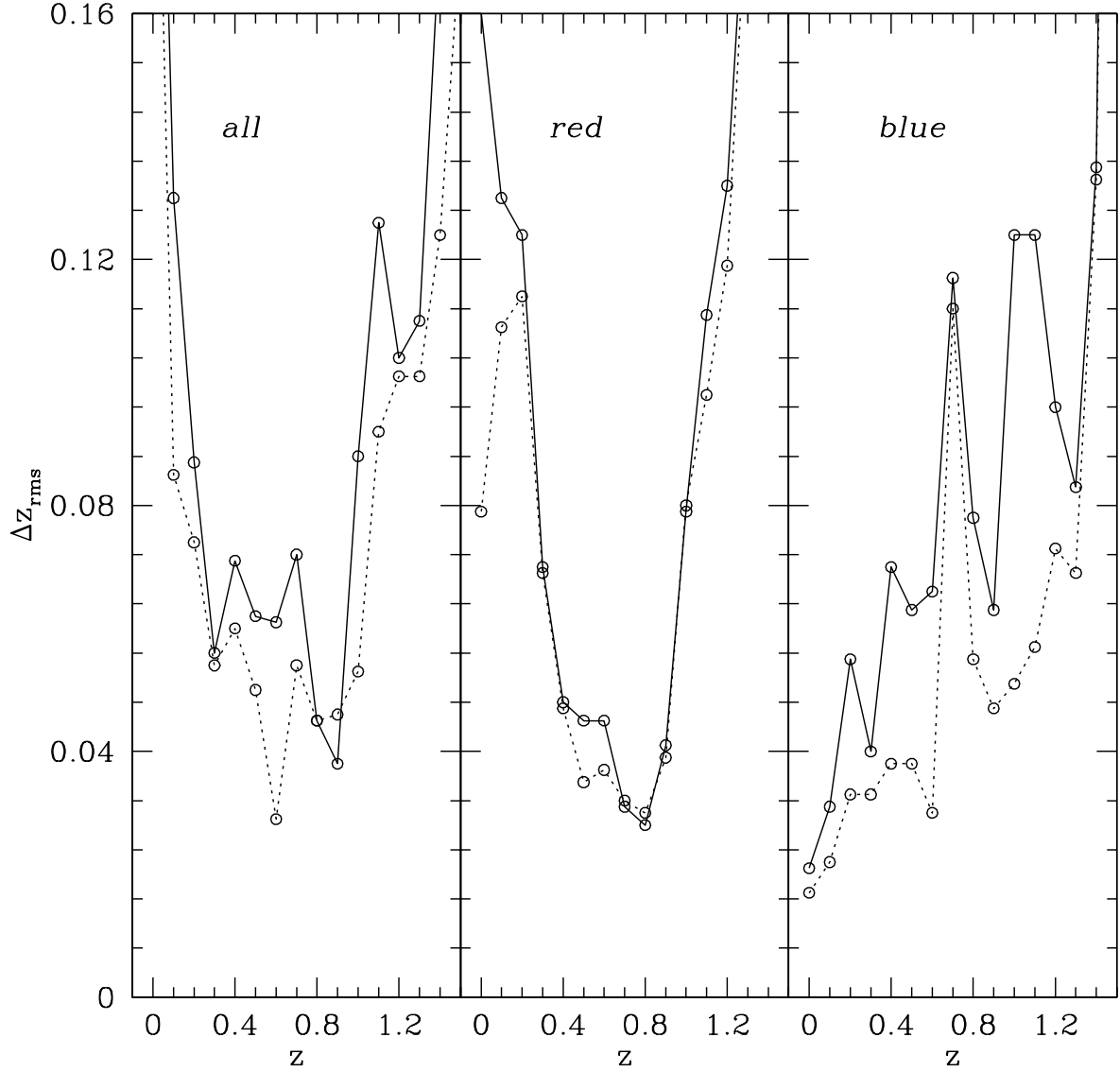}
\caption[Comparison of photometric redshift accuracy]{The rms error on
photometric redshifts using the original filter set ({\it solid line}) and
the extended set ({\it dotted line}).  {\it Left to right}: The improvement
for the entire population in the simulation and for the red (E/S0) and blue
subsamples, respectively.  The large errors at low redshifts are due to the
lack of the $U$ band; at redshift $z>1$, one would need a $J$ band or higher.}
\label{fig:multi.figrms}
\end{figure}

Figure~\ref{fig:multi.figrms}  compares the performance  of the  original and
extended filter sets by plotting the  rms error as a function of
redshift.  The  accuracy of  photometric redshifts for  the entire  catalog is
shown first.  The prediction  improves at all redshifts; the overall
improvement is
\begin{equation}
{\cal Q} = \frac{\sigma_{\rm orig}^2-\sigma_{\rm ext}^2}{\sigma_{\rm orig}^2} 
	 = 34\%.
\end{equation} 
Early- and late-type subsamples are shown in the middle and right panels of
Figure~\ref{fig:multi.figrms}, respectively.  The E/S0 estimates clearly get
better at the smallest redshifts, $z \gtrsim 0$, and at around $z \lesssim
1.2$, when the 4000\,{\AA} break is in one of the humps.  However, the
redshift prediction also improves in another redshift region, around $z \sim
0.55$, as a result of the H\,$\beta$ and Mg\,{\sc i} absorption as it passes
through the second hump.  In according with our expectations, redshift
estimates of the late-type galaxies change much more drastically then those
estimated for early types.

The large number of parameters makes it difficult to estimate the errors or
covariances.  In Table~\ref{tab:multi.tbl}, the rms errors computed for
different set of filters are shown.  The original ACS filter set gives
$\sigma_z = 0.075$. If we try additional multihump filters with humps
centered on the {\it g'r'i'} and {\it g'r'z'} bands, the performance just
slightly improves; $\sigma_z=0.072$ and $0.073$, respectively.  Centering
humps on the gaps between the original bands (off-band multihump filter) does
not improve on the accuracy of the estimates either; $\sigma_z = 0.073$.  The
optimized filter yields much better redshifts, $\sigma_z=0.058$.  The gain in
the entire catalog is comparable to what would be achievable by using
separate IM filters.  The redshift accuracy can be marginally increased by
splitting the optimal multihump into two separate filters ($\sigma_z=0.055$),
but the additional exposure time would be more than twice as long as in our
approach.

\section{Conclusion} \label{sec:multi.conc}

We propose the use of multihump filters extending standard photometric
systems to make photometric redshift estimation much more efficient.  Our
simulations have shown that an optimized extra filter could improve the
redshift prediction by 34\% for the Advanced Camera for Surveys.  The
improvement takes place over the $z<1.3$ redshift range. The filter's
effective width is about 1500\,\AA, comparable to the existing photometric
bands, so the measurement would require only about 25\% more exposure time.
Such a filter can be built with today's technology; in fact, similar filters
have been utilized to suppres atmospheric features at near-infrared
wavelengths \citep{offer98}.  The shape of the extra filter can be tuned to
different goals and missions.  It can be optimized for different filter sets
and adjusted to desired redshift ranges.

\acknowledgements 

I. C. and T. B. acknowledge partial support from the Hungarian Academy of
Sciences--NSF grant 124 and the Hungarian National Scientific Research
Foundation grant T030836, A. J. C. acknowledges support from NASA through a
Long Term Space Astrophysics grant (NAG 5-7934) and through grant
GO-07817.06-96A from the Space Telescope Science Institute.  A. S.
acknowledges support from the NSF (AST 98-02980) and the NASA Long Term Space
Astrophysics program (NAG 5-3503).

\newpage

\begin{deluxetable}{lccccccccc}
\tablecolumns{8}
\tablewidth{0pc}
\tablecaption{\sc Errors on Photometric Redshifts}
\tablehead{
\colhead{Filter Set} & \colhead{$\Delta z_{\rm rms}$} &
\colhead{$\lambda_{\rm start}$} & \colhead{$\lambda_{\rm end}$} &
\colhead{$\lambda_1$} & 
\colhead{$\Delta\lambda_1$} &
\colhead{$\lambda_2$} & 
\colhead{$\Delta\lambda_2$} &
\colhead{$p$}&
\colhead{$\epsilon$}}
\startdata
Original  & 0.075 & \mbox{...} & \mbox{...} & \mbox{...} & \mbox{...} & \mbox{...} & \mbox{...} & \mbox{...} & \mbox{...} \\
{\it g'r'i'} & 0.072 & 4800 & 7900 & 5700 & ~450 & 7000 & 600 & ~8 & 0.05 \\
{\it g'r'z'} & 0.073 & 4900 & 8900 & 5700 & ~450 & 7550 & 900 & ~8 & 0.05 \\
Off-band &     0.073 & 5200 & 8800 & 6200 & ~400 & 7800 & 400 & ~8 & 0.05 \\
Optimized & 0.058 & 4000 & 8500 & 6250 & 1750 & \mbox{...} & \mbox{...} & 20 & 0.05\\
Separate  & 0.055 & \mbox{...} & \mbox{...} & \mbox{...} & \mbox{...} & \mbox{...} & \mbox{...} & \mbox{...} & \mbox{...}\\
\enddata
\label{tab:multi.tbl}
\end{deluxetable}

\end{document}